\documentclass[reprint,superscriptaddress,nofootinbib,amsmath,amssymb, amsfonts,floatfix,aps,prc]{revtex4-2}
\usepackage{subfigure}%
\usepackage{isotope}
\usepackage{siunitx}
\usepackage{graphicx}
\usepackage{bm}
\usepackage{amscd}
\usepackage{hyperref}

\graphicspath{{figs/}}

\newcommand{\ve}[1]{\ensuremath{\mathbf{#1}}}

\makeatletter
\newcommand{\n}[1]{\ensuremath{|%
  \def\splitSubscript##1_##2{\mathbf{##1}_{##2}}%
  \ifx\relax\detokenize\expandafter{\romannumeral-`q##1}\relax
    \mathbf{#1}% Fallback if no subscript
  \else
    \contains@subscript#1__\\{#1}%
  \fi|}}

\def\contains@subscript#1_#2_#3\\#4{%
  \ifx\relax#2\relax
    \mathbf{#4}% No subscript present
  \else
    \splitSubscript#4% Subscript present, split it
  \fi}
\makeatother

\newcommand{\real}{\operatorname{Re}}

\newcommand{\etal}{{\it et al.}}

\newcommand{\epol}{\ensuremath{\vec{e}^{\,+}} }

\begin{document}

\title{Probing the nucleon axial form factor via the $ \vec{e}^{\,+} + \isotope[2]{H} \rightarrow \bar{\nu}_e + p + p$ reaction\\ below the pion-production threshold}

\author{A. M. Ankowski}
\affiliation{M. Smoluchowski Institute of Physics,\\ Faculty of Physics, Astronomy, and Applied Computer Science,\\ Jagiellonian University, PL-30348 Krak\'ow, Poland}
\author{J. Golak}
\affiliation{M. Smoluchowski Institute of Physics,\\ Faculty of Physics, Astronomy, and Applied Computer Science,\\ Jagiellonian University, PL-30348 Krak\'ow, Poland}
\author{E. P\'erez del Rio}
\affiliation{M. Smoluchowski Institute of Physics,\\ Faculty of Physics, Astronomy, and Applied Computer Science,\\ Jagiellonian University, PL-30348 Krak\'ow, Poland}
\author{S. Sharma}
\affiliation{M. Smoluchowski Institute of Physics,\\ Faculty of Physics, Astronomy, and Applied Computer Science,\\ Jagiellonian University, PL-30348 Krak\'ow, Poland}
\author{R.~Skibi{\'n}ski}
\affiliation{M. Smoluchowski Institute of Physics,\\ Faculty of Physics, Astronomy, and Applied Computer Science,\\ Jagiellonian University, PL-30348 Krak\'ow, Poland}
\author{K. Topolnicki}
\affiliation{M. Smoluchowski Institute of Physics,\\ Faculty of Physics, Astronomy, and Applied Computer Science,\\ Jagiellonian University, PL-30348 Krak\'ow, Poland}
\author{H. Wita{\l}a}
\affiliation{M. Smoluchowski Institute of Physics,\\ Faculty of Physics, Astronomy, and Applied Computer Science,\\ Jagiellonian University, PL-30348 Krak\'ow, Poland}
\author{W.~N. Polyzou}
\affiliation{Department of Physics and Astronomy, The University of Iowa, Iowa City, Iowa 52242, USA}
\author{H. Kamada}
\affiliation{Research Center for Nuclear Physics, Osaka University,\\ Ibaraki 567-0047, Japan, and Department of Physics, Faculty of Engineering,
Kyushu Institute of Technology, Kitakyushu 804-8550, Japan}
\author{D. Dutta}
\affiliation{Department of Physics and Astronomy, Mississippi State University, Starkville, MS 39762, USA}

%Collaboration name if desired (requires use of superscriptaddress
%option in \documentclass). \noaffiliation is required (may also be
%used with the \author command).
%\collaboration can be followed by \email, \homepage, \thanks as well.
%\collaboration{}
%\noaffiliation

\date{\today}

%\begin{abstract}
%We study the positron-capture reaction on the deuteron, $\vec{e}^{\,+} + \isotope[2]{H} \rightarrow \bar{\nu}_e + p + p$, for fully longitudinally polarized positron beams at energies up to 150 MeV. Employing a relativistic framework in momentum space, we provide predictions for the total cross section. We also investigate the sensitivity of various differential cross sections to the uncertainty in a~state-of-the-art parametrization of the nucleon axial form factor, the reduction of which is of great importance for long-baseline neutrino-oscillation experiments. Our results are intended to inform future experimental programs at facilities such as Jefferson Lab.  
%\end{abstract}

\begin{abstract}
The nucleon axial form factor $F_A(Q^2)$ is a primary source of systematic uncertainty in charged-current quasielastic interactions, of critical importance for long-baseline neutrino-oscillation studies. Its low-$Q^2$ behavior is particularly problematic to pin down in neutrino measurements. To address this limitation, we investigate the weak process of polarized positron capture on the deuteron, $\vec{e}^{\,+} + \isotope[2]{H} \rightarrow \bar{\nu}_e + p + p$. Employing a relativistic momentum-space formalism, we present predictions for the total cross section. Focusing on the kinematic regime free from pion-production backgrounds, we analyze various differential cross sections and their sensitivity to the $F_A(Q^2)$ variation. Our results demonstrate that thanks to the positron upgrade, Jefferson Lab will gain a unique position to make inroads into determining the low-$Q^2$ behavior of the axial form factor.
\end{abstract}

% insert suggested PACS numbers in braces on next line
%\pacs{23.40.-s, 21.45.-v, 27.10.+h}
% insert suggested keywords - APS authors don't need to do this
%\keywords{}

%\maketitle must follow title, authors, abstract, \pacs, and \keywords
\maketitle

% body of paper here - Use proper section commands
% References should be done using the \cite, \ref, and \label commands

\section{Introduction}
\label{sec:introduction}
%\newpage
Even before the construction of Jefferson Lab commenced~\cite{Westfall:2019}, Hwang proposed~\cite{Hwang:1986nz} to utilize its potential positron beam to perform an ambitious measurement of the nucleon axial form factor, $F_A$, via polarized positron capture on the deuteron, $\epol + \isotope[2]{H} \rightarrow \bar{\nu}_e + p + p$.  The author of Ref.~\cite{Hwang:1986nz} argued that from a theoretical perspective, a measurement of $F_A$ is as important as that of the neutron electric form factor, and that the functional dependence of $F_A$ on the four-momentum transfer squared, $Q^2$, is poorly determined from neutrino scattering on the deuteron.

These arguments are even more compelling today. The dipole parametrization of the axial form factor---considered in Ref.~\cite{Hwang:1986nz} and widely employed until recently\hspace{0pt}---is currently deemed insufficient to capture its $Q^2$ dependence in view of both lattice QCD~\cite{Jang:2023zts} and experimental~\cite{MINERvA:2023avz} findings. In particular, lattice QCD indicates different behavior both at high $Q^2$ and near $Q^2=0$. 

The $F_A(Q^2)$ extractions from neutrino-deuteron scattering experiments performed in the 1980s and early 1990s~\cite{Baker:1981su,Miller:1982qi,Kitagaki:1983px,Kitagaki:1990vs,Allasia:1990uy} have recently been found inconsistent~\cite{MINERvA:2025ygc} with the state-of-the-art MINERvA result obtained from an analysis of antineutrino-hydrogen interactions statistically selected from events collected on a~plastic scintillator target~\cite{MINERvA:2023avz}.

Besides the modeling of nuclear effects, the axial form factor is currently the dominant source of systematic uncertainty in neutrino charged-current quasielastic (CCQE) interactions with atomic nuclei. To achieve its full sensitivity, the Deep Underground Neutrino Experiment (DUNE) requires the total cross sections to be known with 1\% precision~\cite{DUNE:2015lol}. For comparison, the MINERvA $F_A(Q^2)$ parametrization yields a~10--20\% uncertainty in the kinematics of DUNE, corresponding to neutrino energies between 0.5 and 5 GeV. As a consequence, improved determination of the axial form factor is among the prime objectives for precision neutrino-oscillation studies.

\begin{figure}[b]
\includegraphics[width=0.8\columnwidth]{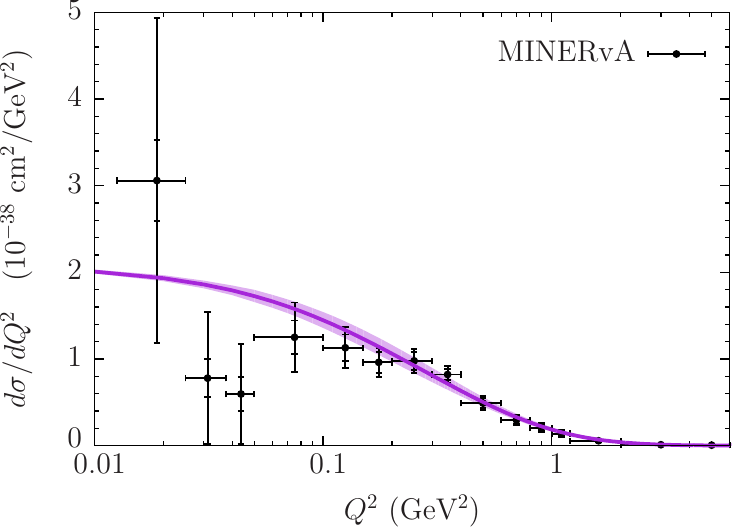}
\caption{\label{fig:MINERvA}Single-differential cross section $d\sigma/dQ^2$ as a function of the four-momentum transfer squared for the process $\bar{\nu}_\mu+p\to\mu^++n$ in the MINERvA experiment. The result calculated using the MINERvA parametrization of the axial form factor is compared with the data from which it was extracted~\cite{MINERvA:2023avz}. The outer (inner) error bars show the total (systematic) uncertainty. The band represents the uncertainty stemming from the form factor.}
\end{figure}

While the value of $F_A(Q^2=0)$ is established from neutron-decay experiments with a few per-mill precision~\cite{Wietfeldt:2023mdb}, uncertainties in the MINERvA data~\cite{MINERvA:2023avz} at low $Q^2$ are particularly large, as shown in Fig.~\ref{fig:MINERvA}. Because the antineutrino energy is unknown, reconstructing $Q^2$ requires the final-state kinematics to be fully determined, where neutron measurement is the limiting factor. 

To contrast this with an ideal target system, the corresponding neutrino CCQE interaction on deuterium, $\nu_\mu + \isotope[2]{H} \to \mu^{-} + p + p$, offers a clear advantage due to the two well-measurable protons in the final state. Currently, however, there are no prospects for such experiments due to the safety concerns associated with handling large volumes of deuterium in underground facilities.

The low-$Q^2$ region plays an important role in neutrino CCQE scattering on complex nuclear targets, such as argon employed in DUNE. For energies of several hundred MeV and higher, $d\sigma/dQ^2$ peaks at $\sim$0.15~GeV$^2$~\cite{Ankowski:2010yh}, resulting in a universal $\sim$20\% contribution of the $Q^2\lesssim0.15$~GeV$^2$ region to the cross section~\cite{Ankowski:2008df}.  

The positron upgrade will place Jefferson Lab in an ideal position to perform groundbreaking studies of the nucleon axial form factor. In particular, measurements at low beam energies---free from pion-production backgrounds---can demonstrate the proof of concept, perform unprecedented tests of low-$Q^2$ behavior, and set new precision standards. 

Here, we employ a relativistic framework validated against electron-scattering data to obtain the total cross section as a function of energy, and analyze the sensitivity of various differential cross sections to the nucleon axial form factor. We also discuss how a measurement of positron capture on the deuteron may be affected by detection thresholds and angular acceptance of the detector.

The paper is organized as follows. In Sec.~\ref{sec:kinematics} we analyze the kinematics of positron capture on the deuteron. In Sec.~\ref{sec:formalism} we present the relativistic momentum-space formalism, developed in Refs.~\cite{PRC98.015501,PRC107.024617} for electron and (anti)neutrino scattering off the deuteron, and apply it to polarized positron capture. The numerical results are presented in Sec.~\ref{sec:results}. Finally, in Sec.~\ref{sec:summary} we summarize our findings and present concluding remarks.

\section{Reaction kinematics}
\label{sec:kinematics}
In the laboratory frame, where the deuteron is at rest, the energy and three-momentum conservation equations read 
\begin{equation}\begin{split}\label{eq:energy&momentumConservation} 
E + M_d &= E' + E_1 + E_2,\\
\ve k &= \ve k' + \ve p_1 + \ve p_2,
\end{split}\end{equation}
where $(E=\sqrt{m_e^2+\ve k^2},\,\ve k)$ and $(E'=\n{k'},\,\ve k')$ denote the energy and momentum of the incident positron and outgoing antineutrino (treated as massless), respectively. The final-state protons have momenta $\ve p_i$ and energies $E_i=\sqrt{M_p^2+\ve p_i^2}$ ($i=1,2$), while $M_d$, $m_e$, and $M_p$ represent the masses of the deuteron, positron, and proton.

The threshold energy of the positron, $E_\text{thr}$, can be determined from the condition on the Mandelstam variable~$s$:
\[ 
\left( E + M_d \right)^2 - \ve k^2 \ge \left(2 M_p\right)^2,
\]
which is obtained assuming a negligible antineutrino mass. Therefore, the reaction can occur when the positron energy exceeds the value
\begin{equation}\label{eq:Ethr}
E_\text{thr} \equiv \frac{ 4 M_p^2 - M_d^2 - m_e^2}{2 M_d}= 931.4\text{ keV},
\end{equation}
which we estimate by taking the $M_d$ value from Ref.~\cite{Wang:2021xhn}.

%We observe that there is no kinematic restriction on the scattering angle $\theta$, , or on the minimal antineutrino energy.

We observe that the antineutrino scattering angle $\theta$, formed by the vectors $\ve k$ and $\ve k'$, is kinematically unrestricted, and that the minimal antineutrino energy always corresponds to its mass, taken to be zero in our calculations.

Let us now obtain the functional dependence of the maximal energy of the antineutrino, $E'_\text{max}$, on the scattering angle $\theta$. The kinematics of the final-state protons is constrained by the condition
\[
(E_1 + E_2)^2-(\ve p_1 + \ve p_2)^2 \ge 4 M_p^2,
\]
which is equivalent to
\[
\left(E + M_d - E'\right)^2 - \left(\ve k - \ve k'\right)^2 \ge 4 M_p^2.
\]
This leads to the following inequality:
\begin{equation}\label{eq:E'max}
	E' \le E'_\text{max}\equiv \frac{ (E + M_d)^2 - \n k^2 - 4 M_p^2 } { 2 \left(E + M_d  - \n k \cos\theta\right) }.
\end{equation}
Therefore, the maximal antineutrino energy is achieved when the final-state protons are at rest in their center-of-mass frame.

Instead of the antineutrino energy, it is convenient to use the energy transfer to the deuteron,
\[\omega \equiv E - E'= E_1 + E_2 - M_d,\]
which is more closely related to the measurable quantities $E_1$ and $E_2$. The maximal antineutrino energy corresponds to the minimal energy transfer, $\omega_\text{min}$.

\begin{figure}
\includegraphics[width=0.8\columnwidth]{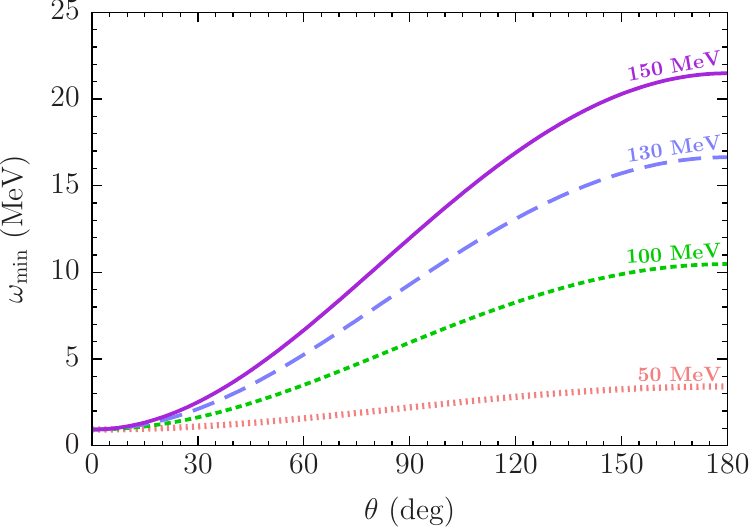}
\caption{\label{fig:minimal_omega}Minimal energy transfer to the deuteron in the positron-capture reaction as a function of the antineutrino scattering angle. Each curve is labeled with its beam energy.}
\end{figure}

The dependence of $\omega_\text{min}$ on the scattering angle is presented in Fig.~\ref{fig:minimal_omega} for positron energies ranging between 50 and 150~MeV. In the limit $\theta\rightarrow0$, the behavior becomes largely universal, with deviations from this picture only of the order of $m_e^2/EM_d$.

% FIG. 2

%To describe the final-state interactions in the two-proton ($2p$) system, we use their invariant mass,
%\begin{equation}\label{eq:E2p}
%W_{2p} \equiv \sqrt{ \left( E + M_d - E' \right)^2 - \ve q^2 },
%\end{equation}
%and their center-of-mass momentum magnitude, $|\ve p_\text{cm}|$, defined via the relation
%\[
%W_{2p} = 2 \sqrt{ M_p^2 + |\ve p_\text{cm}|^2 }.
%\]
%Here, the direction of $\ve p_\text{cm}$ is chosen to correspond to the momentum of the nucleon moving in the forward hemisphere relative to the momentum transfer,
%\[ \ve q \equiv \ve k - \ve k'=\ve p_1 + \ve p_2.\]

To describe the final-state interactions in the two-proton ($2p$) system, we use their invariant mass,
\begin{equation}\label{eq:E2p}
W_{2p} \equiv \sqrt{ \left( E + M_d - E' \right)^2 - \ve q^2 },
\end{equation}
and the magnitude of their momentum in the center-of-momentum (CM) frame, $|\ve p_\text{cm}|$, defined via the relation
\begin{equation}\label{eq:W2p}
W_{2p} = 2 \sqrt{ M_p^2 + |\ve p_\text{cm}|^2 }.
\end{equation}
In Eq.~\eqref{eq:E2p}, $\ve q$ denotes the momentum transfer,
\[ \ve q \equiv \ve k - \ve k'=\ve p_1 + \ve p_2.\]

The proton momenta $\ve p_1$ and $\ve p_2$ in the laboratory frame can be expressed in terms of $\ve p_\text{cm}$ through the Lorentz transformation:
\begin{equation}\begin{split}\label{eq:p1p2} 
\ve{p}_1 & =  \left[\frac12 + \frac{\ve p_\text{cm} \cdot \ve{q}}{W_{2p} ( W_{2p} + E_{2p} ) }  \right] \ve{q} + \ve p_\text{cm},\\
\ve{p}_2 & =  \left[\frac12 - \frac{\ve p_\text{cm} \cdot \ve{q}}{W_{2p} ( W_{2p} + E_{2p} ) }  \right]\ve{q} -\ve p_\text{cm},
\end{split}\end{equation}
where $E_{2p}=E_1 + E_2$ is the total energy of the $2p$ system in the lab frame.

Next, let us determine the kinematic region that is safe from pion-production backgrounds. The primary weak background channel is neutral pion production, ${e^+} + \isotope[2]{H} \rightarrow \bar{\nu}_e + p + p + \pi^0$. Equation~\eqref{eq:E2p} shows that for a given positron energy $E$, the maximal invariant mass of the two-proton system, $W_{2p}^\text{max}$, is reached when the antineutrino is at rest:
\[
W_{2p}^\text{max} = \sqrt{2 E M_d + M_d^2 + m_e^2}.
\]  
This pion-production channel is kinematically disallowed when $W_{2p}^\text{max} < 2 M_p + m_{\pi^0}$, with $m_{\pi^0}$ being the $\pi^0$ mass, which constrains the positron energy to  
\begin{equation} 
	E < \frac{(2 M_p+m_{\pi^0})^2-M_d^2-m_e^2}{2M_d}.
\label{Epi0thr}
\end{equation}
Evaluating this condition yields $E < 140.83$~MeV. For completeness, we note that at higher energies, the electromagnetic process $e^+ + \isotope[2]{H} \rightarrow e^+ + p + p + \pi^-$ also constitutes a background if the final-state positron and pion escape detection; the threshold for this channel is $E = 146.31$~MeV.

We now have the ingredients necessary to consider the range of $Q^2 \equiv \ve q^2 - \omega^2$ accessible in an experiment studying positron capture on the deuteron. This quantity can be expressed as
\[
Q^2 = Q^2(\cos\theta) = 2(E - \sqrt{E^2-m_e^2}\cos\theta)E' - m_e^2.
\]
Recall that the antineutrino's energy (in the massless approximation) can vary between $0 \le E' \le E'_\text{max}$ for $0 \le \theta \le \pi$. For a given $E'$ value, the $Q^2$ range is 
\[
Q^2(\cos\theta=+1) \le Q^2 \le Q^2(\cos\theta=-1).
\]
These two boundaries are increasing functions of $E'$. As a consequence, for a given beam energy, the lower boundary corresponds to the antineutrino at rest, and the upper one to the maximal antineutrino energy:
\begin{equation}
-m_e^2 \le Q^2 \le 2(E + \sqrt{E^2-m_e^2})E'_\text{max} - m_e^2,
\label{eq:Q2limits}
\end{equation}
with $E'_\text{max}$ given in Eq.~\eqref{eq:E'max}.

% FIG. 3
\begin{figure}
\includegraphics[width=0.8\columnwidth]{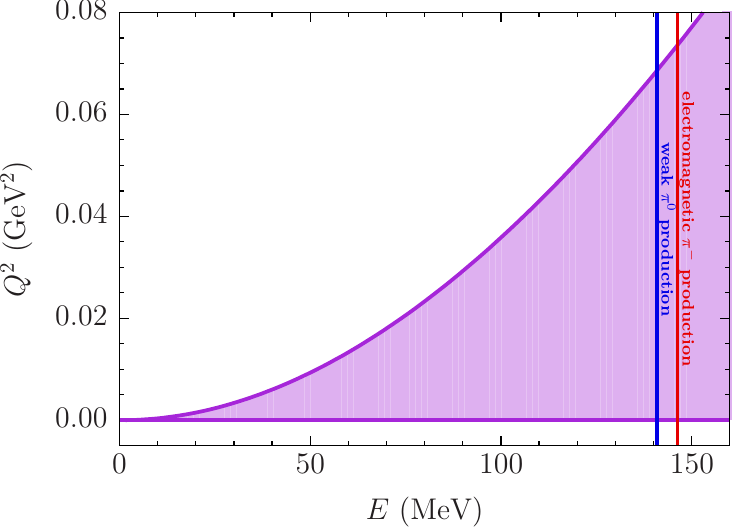}
\caption{\label{fig:Q2range}Kinematically allowed $Q^2$ range as a function of the positron energy. The vertical lines show the thresholds for the $e^+ + \isotope[2]{H} \rightarrow \bar\nu_e + p + p + \pi^0$ and $e^+ + \isotope[2]{H} \rightarrow e^+ + p + p + \pi^-$ background channels. }
\end{figure}

Figure~\ref{fig:Q2range} presents the allowed $Q^2$ range for beam energies ranging from the threshold~\eqref{eq:Ethr} up to 160~MeV. For reference, we also indicate the thresholds for weak $\pi^0$ and electromagnetic $\pi^-$ production, which constitute backgrounds for the positron-capture process. For $E=130$~MeV, safe from the pion backgrounds, the maximal $Q^2$ value is $\sim$0.059 GeV$^2$.  

\section{Relativistic Momentum-Space Formalism}
\label{sec:formalism}

In the process of positron capture on the deuteron, the element of the differential cross section in a general frame can be expressed as
\begin{equation}\begin{split}\label{eq:crossSectionElement_generalFrame}
d \sigma & = \frac12\,\frac{(2\pi)^4}{v_\text{rel}}\delta^4(k + p_d - k' - p_1 - p_2)\\
&\quad\times |\langle \ve p_1, \ve p_2, \ve k' \Vert T \Vert \ve p_d, \ve k \rangle |^2 \, d^3\ve k' \, d^3\ve p_1 \, d^3\ve p_2,
\end{split}\end{equation}
where the statistical factor $\frac12$ appears due to two identical particles in the final state, $v_\text{rel}$ denotes the relative speed of the projectile and the target,
\[
v_\text{rel} = \frac{\sqrt{(k\cdot p_d)^2 - M_d^2 m_e^2}}{E E_d},
\]
and $p_d = (E_d, \ve p_d)$ is the four-momentum of the deuteron. We suppress spin and isospin quantum numbers, writing the transition matrix stripped of the three-momentum-conserving delta function simply as $\langle \ve p_1, \ve p_2, \ve k' \Vert T \Vert \ve p_d, \ve k \rangle$. 

\subsection{Gell-Mann--Goldberger approach}
For weak scattering, the transition operator can be approximated using the Gell-Mann--Goldberger two-\hspace{0pt}potential formalism~\cite{gell-mann-goldberger}, where the strong interaction is treated exactly and the weak interaction is accounted for perturbatively at leading order. Then, the dynamical unitary representation of the Poincar\'e group factors into a tensor product of strong and weak representations:
\[
U(\Lambda, a) \to U_\text{str} (\Lambda, a) \otimes U_\text{weak}(\Lambda, a),
\]
and the reduced transition matrix elements take the form
\[
\langle \ve p_1, \ve p_2, \ve k' \Vert T \Vert \ve p_d, \ve k \rangle \approx \frac{G_F \cos\theta_C}{\sqrt{2}}\eta_{\mu\nu} N^{\mu}\langle \ve k'|J^{\nu}_\text{weak}(0)|\ve k \rangle,
\]
where $J^{\nu}_\text{weak}(0)$ is the weak current evaluated at the spacetime origin, $G_F$ is the Fermi constant, and $\theta_C$ is the Cabibbo angle. In our convention, the metric tensor is $\eta_{\mu\nu}=\mathrm{diag}(1,-1,-1,-1)$. The strong-current matrix element,
\begin{equation}
N^{\mu} \equiv \langle (\ve p_1, \ve p_2)^+ | J_\text{str}^{\mu} (0)| \ve p_d \rangle,
\end{equation}
%describes the transition between the deuteron state with momentum $\ve p_d$ and a two-nucleon scattering state.
describes the transition between the deuteron state with momentum $\ve p_d$ and the exact two-nucleon outgoing scattering state $\vert{}(\ve p_1, \ve p_2)^+\rangle = \Omega_+ \vert{}\ve p_1, \ve p_2\rangle$.

The weak-current matrix element associated with the transition of a positron to an antineutrino,
\[
\langle \ve k'|J^{\mu}_\text{weak}(0)|\ve k \rangle = \frac{1}{(2 \pi)^3} \frac{1}{2\sqrt{E E'}} L^{\mu},
\]
with
\begin{equation}\label{eq:L}
L^\mu = \bar{v}(\ve k, s_{e^+}) \gamma^\mu (1 - \gamma_5) v(\ve k', s_{\bar{\nu}_e}),
\end{equation}
depends on the positron's three-momentum $\ve k$ and spin projection $s_{e^+}$, as well as on the antineutrino's three-momentum $\ve k'$ and spin projection $s_{\bar{\nu}_e}$. We adopt here the Bjor{\-}ken and Drell~\cite{Bjorken} normalization for the Dirac spinors, $\bar{v}(\ve k, s)\, v(\ve k,s') = -\delta_{ss'}$.

The element of the differential cross section~\eqref{eq:crossSectionElement_generalFrame} then becomes
\begin{equation}\begin{split}\label{eq:crossSectionElement_IA}
d \sigma & = \frac{G_F^2 \cos^2\theta_C}{16(2\pi)^2 v_\text{rel} EE'}\,\delta^4(k + p_d - k' - p_1 - p_2)\\
&\quad\times L^\mu (L^\nu)^* N_{\mu} N_{\nu}^* \, d^3\ve k' \, d^3\ve p_1 \, d^3\ve p_2.
\end{split}\end{equation}
In numerical calculations, we use $G_F = 1.1663785 \times 10^{-5}~\text{GeV}^{-2}$ and $\cos\theta_C = 0.97367$~\cite{ParticleDataGroup:2026}. We determine the wave functions describing the deuteron bound state using the proton-neutron version of the Argonne $v_{18}$ potential~\cite{AV18}.

\subsection{Impulse approximation}
\label{sec:formalism.IA}
Due to the covariance requirement, both one- and two-body terms appear in the nuclear current. Unlike one-body currents, the two-body terms are representation-dependent. Their role in (anti)neutrino interactions with the deuteron was investigated, for example, in Refs.~\cite{Marcucci11,PRC86.035503}, where they were found to contribute no more than 4\% to the total cross section at the considered kinematics. We will return to this point when discussing our results in Sec.~\ref{sec:results}.

Therefore, we employ the impulse approximation, that is, we confine our analysis to single-nucleon currents. These describe the weak charged-current neutron-to-proton transition and can be parametrized in the standard form~\cite{LlewellynSmith:1971zm,Vogel:1999zy,Strumia:2003zx}:
\[\begin{aligned}
& \big\langle \ve p', s', \tau'= \tfrac12 \,\big|\, j_\text{CC}^\mu(0) \,\big|\, \ve p, s, \tau=-\tfrac12 \big\rangle\\
&\quad= \bar u(\ve p', s') \bigg( \gamma^\mu F_1 + i\sigma^{\mu\nu}\frac{q_\nu}{2 M_N} F_2 \\
&\qquad\qquad\qquad+ \gamma^\mu \gamma_5 F_A + \frac{q^\mu}{M_N}\gamma_5 F_P \bigg) u(\ve p, s),
\end{aligned}\]
where $M_N$ denotes the average nucleon mass, and the weak form factors $F_x$ ($x = 1, 2, A, P$) are functions of $Q^2$, parametrized according to Refs.~\cite{Bradford:2006yz,MINERvA:2023avz}. 

Because the vector form factors $F_1$ and $F_2$ are related to the electromagnetic form factors of the proton and neutron~\cite{Gershtein:1955fb,Feynman:1958ty} and the pseudoscalar form factor $F_P$ can be expressed in terms of the axial form factor~\cite{Goldberger:1958vp}, only $F_A$ is a relevant source of uncertainty in positron capture.

\subsection{Poincar\'e dynamics of two-nucleon states}
In evaluating the nuclear current matrix element $N^\mu$, the wave functions of the deuteron target and the final $2p$ continuum state are naturally determined in the laboratory and CM frames, respectively. This difference makes it essential to formulate their relativistic transformations explicitly.

In Dirac's instant form of dynamics~\cite{Dirac:1949,relform1,kei91}, boost generators explicitly depend on the nuclear interaction, making this frame transformation dynamical rather than purely kinematic. The relativistic transformations of both bound and scattering states are thus governed by a unitary representation, $U(\Lambda, a)$, of the Poincar\'e group acting on the two-nucleon Hilbert space, ensuring that all physical observables---such as probabilities, expectation values, and ensemble averages---are frame-independent.

The model Hilbert space is the tensor product of two single-particle spaces. A basis for a single particle of mass $m$ and spin $s = 1/2$ consists of simultaneous eigenstates of momentum and projections of spin and isospin,
\[
\vert (m, s) \ve p, \mu, \tau \rangle.
\]
where $(m, s)$ label the invariant mass and spin of the single-particle Poincar\'e representation, while $\ve p$, $\mu$, and $\tau$ denote the state quantum numbers.

In this basis, the single-particle unitary representation of the Poincar\'e group acts as
\[\begin{split}
&U_{ms}(\Lambda, a) \lvert (m, s) \ve p, \mu, \tau \rangle \\
&\qquad = \sum_{\nu=-1/2}^{1/2} e^{-i \Lambda p\cdot a} \lvert (m, s) \boldsymbol{\Lambda}\ve p, \nu, \tau \rangle \\ &\qquad\qquad \times \sqrt{ \frac{E(\boldsymbol{\Lambda}\ve p)}{E(\ve p)} }\,D^{s}_{\nu\mu} \!\left[ B^{-1}(\boldsymbol{\Lambda}\ve p/m) \Lambda B(\ve p/m) \right],
\end{split}\]
where $\Lambda$ is a Lorentz transformation, $a$ denotes a spacetime displacement, $E(\ve p) = \sqrt{m^2 + \ve{p}^2}$, and $B(\ve p/m)$ is a pure boost transforming a particle at rest $(m, \ve 0)$ to momentum $\ve p$. The argument of the Wigner $D$-matrix represents the standard Wigner rotation, and the momentum eigenstates are normalized as $\langle\ve p'\mid \ve p\rangle = \delta^3(\ve p - \ve p')$.

The dynamics of two \emph{free} nucleons is given by the tensor product of the single-nucleon unitary representations of the Poin{\-}car\'e group,
\[
U_0(\Lambda, a) \equiv U_{m_1 s_1}(\Lambda, a) \otimes U_{m_2 s_2}(\Lambda, a).
\]
The tensor product representation is reducible, as states with different invariant masses and spins do not mix. Using the Clebsch-Gordan coefficients of the Poincar\'e group given in Ref.~\cite{PRC107.024617}, the tensor product states can be decomposed into a linear superposition of simultaneous eigenstates of invariant mass $m_{12}(\kappa)$, total spin $s$, total momentum $\ve p$, magnetic quantum number $\mu$, and total isospin projection $\tau_{12}$. The resulting irreducible basis states take the form
\[
\lvert (\kappa, s) \ve p, \mu, \tau_{12} (l_{12}, s_{12}) \rangle,
\]
where $l_{12}$ and $s_{12}$ are invariant degeneracy parameters. The spectra of $l_{12}$ and $s_{12}$ are identical to those of the relative orbital angular momentum and total spin in a standard partial-wave basis. However, the Poincar\'e Clebsch-Gordan coefficients also include momentum-dependent Jacobians and Wigner rotations.

The magnitude parameter $\kappa$ characterizes the invariant mass $m_{12}(\kappa) = \sqrt{m_1^2 + \kappa^2} + \sqrt{m_2^2 + \kappa^2}$ and represents the magnitude of the momentum of either nucleon in the CM frame. When $m_1 = m_2 = M_p$, as in Eq.~\eqref{eq:W2p}, $\kappa = |\ve p_\text{cm}|$ and $m_{12} = W_{2p}$.

The interacting representation of the Poincar\'e group is constructed by embedding an interaction into the free two-body invariant mass operator. In the Bakamjian--Thomas form~\cite{BakamjianThomas1953}, the mass operator $\hat M$ is defined as
\begin{equation}\label{eq:M_mass_op}
\hat M \equiv \sqrt{m_1^2 + \kappa^2 + 2 \mu_\text{red}\hat V} + \sqrt{m_2^2 + \kappa^2 + 2 \mu_\text{red}\hat V},
\end{equation}
where $\mu_\text{red} = m_1 m_2 / (m_1 + m_2)$ is the two-nucleon reduced mass. In the irreducible basis, the matrix elements of the potential operator $\hat V$ take the diagonal form
\[\begin{aligned}
&\langle (\kappa, s) \ve p, \mu, \tau_{12} (l_{12}, s_{12}) \vert \hat V \vert (\kappa', s') \ve p', \mu', \tau_{12}' (l_{12}', s_{12}') \rangle \\
&= \delta^{(3)}(\ve p - \ve p') \delta_{s s'} \delta_{\mu \mu'} \langle \kappa, l_{12}, s_{12}, \tau_{12} \Vert \hat V \Vert \kappa', l_{12}', s_{12}', \tau_{12}' \rangle,
\end{aligned}\]
where $\langle \kappa, l_{12}, s_{12}, \tau_{12} \Vert \hat V \Vert \kappa', l_{12}', s_{12}', \tau_{12}' \rangle$ is the reduced matrix element of the realistic $N\!N$ interaction. We perform calculations using the Argonne $v_{18}$ potential~\cite{AV18}.

It is important to note that the dynamical mass operator $\hat M$ is a monotonic function of the nonrelativistic internal Hamiltonian, $h_\text{NR} = {\kappa^2}/{2\mu_\text{red}} + \hat V$. As a consequence, $\hat M$ and $h_\text{NR}$ share identical internal wave functions and scattering phase shifts as functions of the CM momentum magnitude $\kappa$.

Realistic $N\!N$ potentials, such as Argonne $v_{18}$, are fitted to scattering data and phase shifts parameterized by the CM momentum $\kappa$, with laboratory kinetic energies converted using relativistic kinematics. Thus, the mass operator in Eq.~\eqref{eq:M_mass_op} yields phase shifts that reproduce the \emph{experimentally determined} Poincar\'e-invariant $N\!N$ phase shifts as functions of the relative momentum $\kappa$, and also gives the correct value of the deuteron binding energy. This construction relies on identifying the magnitude of a single particle's momentum in the nonrelativistic CM frame with its counterpart in the relativistic rest frame.

Simultaneous eigenstates of $\hat M$, $s$, $\ve p$, $\mu$, and $\tau_{12}$ exhibit the same Poincar\'e transformation properties as a free state of invariant mass $m_{12}(\kappa)$ and spin $s$, with the free mass eigenvalue replaced by the eigenvalue $M$ of the interacting mass operator $\hat M$. Therefore, it is possible to construct exact Lorentz boosts of the final two-nucleon scattering state to account for the momentum transferred by the probe. Because these boost transformations depend on the eigenvalues of the interacting mass operator $\hat M$ rather than the free mass operator $\hat M_0$---whose eigenvalues are $m_{12}(\kappa)$---they are inherently dynamical.

In this work, instead of Eq.~\eqref{eq:M_mass_op}, we use an $S$-matrix-equivalent mass operator of the form 
\begin{equation}
\hat M' = \sqrt{m_1^2 + \kappa^2} + \sqrt{m_2^2 + \kappa^2} + \hat W_\text{eff},
\label{eq:M_prime}
\end{equation}
where the effective interaction $\hat W_\text{eff}$ is derived from $\hat M$ following the method of Refs.~\cite{KAMADA2007119,Kamada:2014dba}. In this manner, the dynamical problem reduces to solving the mass eigenvalue equation in the irreducible noninteracting basis. The full eigenstates are expanded as
\[\begin{aligned}
&\langle (\kappa, s) \ve p, \mu, \tau_{12} (l_{12}, s_{12}) \vert (M_I, s') \ve p', \mu', \tau'_{12}, d \rangle\\
& = \delta^3(\ve p - \ve p') \delta_{s s'} \delta_{\mu \mu'} \delta_{\tau_{12} \tau'_{12}} \langle \kappa, l_{12}, s_{12}, \tau_{12} \vert \psi_d \rangle,
\end{aligned}\]
where $d$ represents the internal degeneracy parameters, and $M_I$ denotes the invariant mass eigenvalue. The wave function $\lvert \psi_d \rangle$ satisfies the mass eigenvalue problem
\[
\hat M' \lvert \psi_d \rangle = \left( \hat M_0 + \hat W_\text{eff} \right) \lvert \psi_d \rangle = M_I \lvert \psi_d \rangle,
\]
with $\hat M_0$ denoting the free mass operator.

The dynamical unitary representation $U(\Lambda, a)$ of the Poincar\'e group acts on these interacting states as
\[\begin{split}
&U(\Lambda, a) \lvert (M_I, s) \ve p, \mu, \tau_{12}, d \rangle \\
&\quad=  \sum_{\nu=-s}^{s} e^{-i \Lambda p \cdot a} \lvert (M_I, s) \boldsymbol{\Lambda}\ve p, \nu, \tau_{12}, d \rangle \\
&\qquad\times \sqrt{\frac{E_I(\boldsymbol{\Lambda} \ve p)}{E_I(\ve p)}} D^s_{\nu \mu}\!\left[ B^{-1}(\boldsymbol{\Lambda} \ve p/M_I) \Lambda B(\ve p/M_I) \right],
\end{split}\]
where $E_I(\ve p) = \sqrt{M_I^2 + \ve p^2}$ and $B(\ve p/M_I)$ denotes a pure $SL(2,\mathbb{C})$ boost.

\subsection{Nuclear matrix element\break and final-state interactions}

The matrix element $N^\mu$ describes the transition induced by the nuclear current operator from the initial deuteron state $\vert \ve p_d,\mu_d, D \rangle$---where $D$ specifies the discrete bound-state channel ($J=1, T=0$)---to the final antisymmetrized two-nucleon scattering state,
\[\begin{aligned}
\lvert \ve p_1, \mu_1, \tau_1, \ve p_2, \mu_2, \tau_2 \rangle_a & = \frac{1}{2} \Bigl( 
  \lvert \ve p_1, \mu_1, \tau_1, \ve p_2, \mu_2, \tau_2 \rangle \\
  &\qquad - \lvert \ve p_2, \mu_2, \tau_2, \ve p_1, \mu_1, \tau_1 \rangle
\Bigr).
\end{aligned}\]

This element naturally decomposes into a sum of plane-wave and rescattering contributions,
\begin{equation}\label{eq:fullMatrixElement} 
N^\mu = N^\mu_\text{PW} + N^\mu_\text{FSI},
\end{equation}
whose detailed derivation is provided in the appendix of Ref.~\cite{PRC107.024617}.

As explained in Sec.~\ref{sec:formalism.IA}, we employ the impulse approximation, $J_\text{str}^\mu = J_\text{IA}^\mu$, where $J_\text{IA}^\mu$ is the sum of the single-nucleon currents. The matrix-element contributions are calculated as
\[
N^\mu_\text{PW} = {}_a\langle \ve p_1, \mu_1, \tau_1, \ve p_2, \mu_2, \tau_2 \lvert J_\text{IA}^\mu(0) \rvert \ve p_d, \mu_d, D \rangle
\]
and
\[\begin{aligned}
N^\mu_\text{FSI} & = {}_a\langle \ve p_1, \mu_1, \tau_1, \ve p_2, \mu_2, \tau_2 \lvert \\
&\qquad t(E_{2p}+i\epsilon; \n q) G_0 (E_{2p}+i\epsilon) J_\text{IA}^\mu(0) \rvert \ve p_d, \mu_d, D \rangle,
\end{aligned}\]
where $G_0 (E_{2p})$ is the relativistic free two-nucleon propagator, with $E_{2p} = E_1 + E_2$ and $\ve q = \ve p_1 + \ve p_2$. The off-shell transition matrix $t(E_{2p}; \n q)$ satisfies the Lippmann-Schwinger equation,
\[
    t(E_{2p}; \n q)  = v(\n q) + t(E_{2p}; \n q) G_0(E_{2p} + i \epsilon) v(\n q),
\]
with the ``boosted'' potential $v(\n q)$ generated from the nonrelativistic nucleon-nucleon potential $\hat V$~\cite{AV18} according to the approach of Ref.~\cite{KAMADA2007119}. Note that $v(\n q)$ reduces to $\hat W_\text{eff}$ in Eq.~\eqref{eq:M_prime} when $\n q = 0$.

To construct $v(\n q)$, the strong proton-proton version of the $v_{18}$ interaction is augmented by a sharply cut-off Coulomb force $V_{RC}$. This approach is justified by the observation~\cite{PRC98.015501} that the matrix elements, when Bessel-transformed to coordinate space, become negligible for $r \ge R_C$. We have verified that for $R_C \ge 20~\text{fm}$, our results are independent of the cut-off value.

Despite their apparent simplicity, the nuclear matrix elements $N^\mu$ are highly non-trivial to construct even for the single-nucleon current operator, as they rely on the choice of non-interacting irreducible states and the resulting Poincar\'e Clebsch-Gordan coefficients~\cite{moussa,Polyzou11}. Crucially, the plane-wave part is evaluated without resorting to a partial-wave decomposition, whereas the rescattering calculation includes all two-proton partial waves with total angular momentum~$j \le 4$.

\subsection{The cross section in the laboratory frame}
We consider the case where the positron is longitudinally polarized, the deuteron is unpolarized, and the polarizations of the outgoing particles are not detected. While this scenario cannot be realized experimentally at present, it should be within the realm of possibility in the near future. To obtain the total cross section, one needs to average over the initial spin projections of the deuteron $s_d$, sum over the final protons' spins $s_1$ and $s_2$, and integrate over all the final momenta. We also formally sum over the antineutrino spin projections $m_{\bar{\nu}}$ by introducing the leptonic tensor,
\[
L^{\mu\nu}=\sum_{m_{\bar\nu}}L^\mu (L^\nu)^*,
\] 
in order to make use of standard Dirac completeness relations. The unphysical wrong-chirality state automatically gives a vanishing contribution due to the chiral projection operator $(1-\gamma_5)$ in Eq.~\eqref{eq:L}. 

As a consequence of Lorentz invariance of the phase space elements $d^3\ve p_i/2E_i$, the integration over the protons' momenta can be carried out using the Dirac delta function:
\[
\delta^4(\dots) \, d^3\ve p_1 \, d^3\ve p_2 =\frac14{\mathcal J}(\n{p_1},\n{p_2})\,|\ve p_\text{cm}|E_{2p}\, d\Omega_\text{cm},
\]
where $d\Omega_\text{cm}=d\cos\theta_\text{cm}\,d\phi_\text{cm}$, with $\theta_\text{cm}$ and $\phi_\text{cm}$ being the polar and azimuthal angles of $\ve p_\text{cm}$, and the Jacobian ${\mathcal J}(\n{p_1},\n{p_2})$ is given by
\[
{\mathcal J}(\n{p_1},\n{p_2})=\frac{4E_1E_2}{W_{2p}E_{2p}}.
\]
The phase space element for antineutrino can be cast in the form
\[
\frac{d^3\ve k'}{E'} = \frac{\n{ k'}^2}{E'}\, d\cos\theta\,d\phi\, d\n{k'}= \n{ k'}\, d\cos\theta\,d\phi\, dE'.
\]
Due to cylindrical symmetry about the beam direction $\ve{k}$, we can---without loss of generality---fix the lepton 
scattering plane such that the azimuthal angle $\phi$ is set to zero. Then, the azimuthal integration reduces to multiplying by a factor of $2\pi$. 

Noting that in the laboratory frame, the relative speed reduces to $v_\text{rel}=\n{k}/E$, and combining the above ingredients, we arrive at the following expression for the 4-fold differential cross section:
\[\begin{split}
    \frac{d^4\sigma}{ d\cos\theta\, dE' d\Omega_\text{cm} }
     &=  \frac{G_F^2 \cos^2\theta_C} {128\pi\n{k}}\, \n{k'}|\ve p_\text{cm}|\,(E + M_d - E') \,\\ 
&\times{\mathcal J}(\n{p_1},\n{p_2})\frac13 \, \sum_{s_1,\,s_2} \sum_{s_d} L^{\mu\nu}N_\mu N_\nu^*.
\end{split}\]
Its integration yields the total cross section:
\[\begin{split}
    \sigma 
     & = \int_{-1}^{+1} d\cos\theta \int_{0}^{E'_\text{max}} dE'\\ 
     &\quad\times\int_0^{2\pi} d\phi_\text{cm} \int_{-1}^{+1} d\cos\theta_\text{cm}\,\frac{d^4\sigma}{d\cos\theta\,dE' d\Omega_\text{cm}}.
\end{split}\]

It is convenient to choose a coordinate system where the $z$ axis is aligned with the 
3-momentum transfer $\ve{q}$, and the $y$ axis is defined along 
$\ve{k} \times \ve{k}'$. In this frame, the components of the initial and final 
positron 3-momenta read:
\begin{align*}
k'_x & =  k_x  =  \frac{\n{k}\n{k'}}{\n{q}}\sin\theta,\\
k'_y & =  k_y  =  0,\\
k_z & = \frac{\n{k}}{\n{q}} (\n{k}  - \n{k'}\cos\theta),\\
k'_z & =\frac{\n{k'}}{\n{q}}(\n{k}\cos \theta -\n{k'}),
\end{align*}
%\begin{align*}
%k'_x & = k_x  = {\n{k}\n{k'}}\sin\theta/{\n{q}},\\
%k'_y & = k_y  = 0,\\
%k_z & = {\n{k}}(\n{k}  - \n{k'}\cos\theta)/{\n{q}},\\
%k'_z & ={\n{k'}}(\n{k}\cos \theta -\n{k'})/{\n{q}},
%\end{align*}
where $\n{q} = \sqrt{ \n{k}^2 +  \n{k'}^2 - 2 \n{k} \n{k'} \cos \theta}$.

% FIG. 4
%\begin{figure*}
%    \subfigure{\label{fig:electrons_a}}
%    \subfigure{\label{fig:electrons_b}}
%\includegraphics[width=0.8\textwidth]{electrons}
%\caption{\label{fig:electrons}Double differential electron-deuteron cross section, $d^2\sigma/d\omega d\Omega$. The results obtained using the plane-wave approximation and the full calculations are compared with the experimental data reported by (a) Quinn \etal{}~\cite{Quinn:1988ua} and (b) Parker \etal{}~\cite{Parker:1986}. The panels are labeled according to the beam energy and scattering angle. Note that Quinn \etal{} found a normalization problem in the Parker \etal{} data.}
%\end{figure*}

\begin{figure*}
    \subfigure{\label{fig:electrons_a}}
    \subfigure{\label{fig:electrons_b}}
\includegraphics[width=0.8\textwidth]{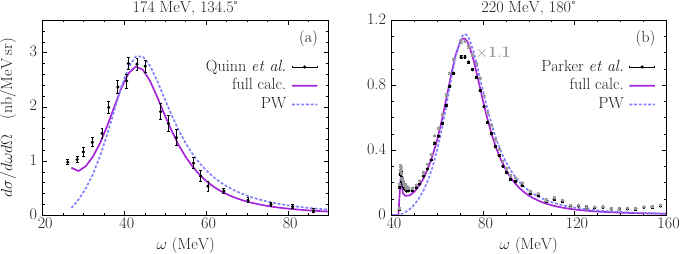}
\caption{\label{fig:electrons}Double-differential electron-deuteron cross section, $d^2\sigma/d\omega d\Omega$. The results obtained using the plane-wave approximation and the full calculation are compared with the experimental data reported by (a) Quinn \etal{}~\cite{Quinn:1988ua} and (b) Parker \etal{}~\cite{Parker:1986}. The panels are labeled according to the beam energy and scattering angle. Because Quinn \etal{} identified a $\sim$10\% normalization discrepancy in the dataset of panel (b), these data are also shown multiplied by 1.1.}
\end{figure*}

The contraction of the leptonic tensor with the nuclear currents,
\[ |\mathcal{T}|^2 \equiv{L^{\mu\nu}}N_\mu N_\nu^*,\]
can be expressed in a spherical basis as
\[\begin{split}
|\mathcal{T}|^2  & = V_{00} |N_0|^2 + V_{mm} |N_{-}|^2 + V_{pp} |N_{+}|^2 + V_{zz} |N_{z}|^2 \\
          &\quad +  2V_{m0}\real \left( N_{-} N_0^* \right) 
           + 2V_{p0}\real\left( N_{+} N_0^* \right) \\
          &\quad +  2V_{z0}\real\left( N_{z} N_0^* \right) 
           +  2V_{mp}\real \left( N_{-} N_{+}^* \right)\\
          &\quad +  2V_{mz}\real\left( N_{-} N_z^* \right) 
           + 2V_{pz}\real\left( N_{+} N_z^* \right).
\end{split}\]
With the mass-weighted helicity spin four-vector of the positron with helicity $h= +1$ defined as 
\[
        \tilde{s}^\alpha\equiv m_e S^\alpha = \left({\n k}, \frac{E} {\n k} k_x, \frac{E} {\n k} k_y, \frac{E} {\n k} k_z\right),
\]
the leptonic coefficients are the following:
\begin{align*}
V_{00} &= V_{zz}\! = 4[ E'(E + \tilde{s}_0)
       - k'_x(k_x + \tilde{s}_x)
       + k'_z(k_z + \tilde{s}_z) ],\\
V_{mm} &= 4\,(E' + k'_z)\,(E - k_z + \tilde{s}_0 - \tilde{s}_z),\\
V_{pp} &= 4\,(E' - k'_z)\,(E + k_z + \tilde{s}_0 + \tilde{s}_z),\\
V_{m0} &= -2\sqrt{2}[(E' + k'_z)(k_x + \tilde{s}_x)\\
&\quad + k'_x(E - k_z + \tilde{s}_0 - \tilde{s}_z)],\\
V_{p0} &= 2\sqrt{2}[(E' - k'_z)(k_x + \tilde{s}_x)\\
&\quad + k'_x(E + k_z + \tilde{s}_0 + \tilde{s}_z)],\\
V_{z0} &= -4[
    k'_z(E + \tilde{s}_0) + E'(k_z + \tilde{s}_z) ],\\
V_{mp} &= -4\,k'_x\,(k_x + \tilde{s}_x),\\
V_{mz} &= 2\sqrt{2}[
   (E' + k'_z)(k_x + \tilde{s}_x)
    - k'_x(E - k_z + \tilde{s}_0 - \tilde{s}_z)],\\
V_{pz} &= 2\sqrt{2}[
   (E' - k'_z)(k_x + \tilde{s}_x)
    - k'_x(E + k_z + \tilde{s}_0 + \tilde{s}_z) ].
\end{align*}

%Numerical results were obtained using a grid of 256 points in $\cos\theta$, 300 points in $E'$, and 5810 $(\theta_\text{cm}, \phi_\text{cm})$ pairs from the Lebedev-Laikov quadrature~\cite{lebedev, Burkardt}, with $\phi = 0$. Each of the $446{,}208{,}000$ points $p(i)$ is characterized by the value of the fivefold differential cross section $d\sigma(i)$ alongside a differential phase-space weight $dW(i)$. The integrated cross section $\Delta\sigma$ over a phase-space region $V$ is evaluated as
%\[
%\Delta\sigma = \sum_{p(i) \in V} dW(i) \, d\sigma(i).
%\]
%This approach allows for straightforward implementation of kinematic cuts tailored to prospective experimental data.

Numerical results were obtained using a grid of 256 points in $\cos\theta$, 300 points in $E'$, and 5810 $(\theta_\text{cm}, \phi_\text{cm})$ pairs from the Lebedev-Laikov quadrature~\cite{lebedev, Burkardt}, with $\phi = 0$. Each of the 446,208,000 points $p(i)$ is characterized by the value of the fivefold-differential cross section $d\sigma(i)$ alongside a differential phase-space weight $dW(i)$.

The volume of a phase-space region $V$ and its contribution to the cross section are evaluated as
\[
|V| = \sum_{p(i) \in V} \, dW(i),\qquad \Delta\sigma = \sum_{p(i) \in V} \, dW(i) \, d\sigma(i),
\]
with the average cross section given by $\langle \sigma \rangle \equiv \Delta\sigma / |V|$.

This approach allows for a straightforward implementation of kinematic cuts tailored to prospective experimental data.

\section{Results}
\label{sec:results}
To make our results directly applicable to experimental conditions, in this section we express all kinematic quantities in the laboratory frame.

Before we apply our framework to positron capture, it is important to validate it against experimental electron-scattering data for the double-differential cross sections collected at similar kinematics for the deuteron. To the best of our knowledge, the lowest-energy datasets available are those extracted by Quinn \etal{}~\cite{Quinn:1988ua} and Parker \etal{}~\cite{Parker:1986} with beam energies of 174 and 220~MeV and scattering angles of $\ang{134.5}$ and $\ang{180}$, respectively.

It is important to note that Quinn \etal{}~\cite{Quinn:1988ua} performed extensive verifications of the normalization of their data---including results for other targets---by comparisons with the results of other experiments and theoretical predictions. Through these comparisons, they found that the cross sections reported by Parker \etal{}~\cite{Parker:1986} were systematically underestimated, showing a $\sim$10\% discrepancy in the 180-MeV dataset.

%\begin{figure}
%    \subfigure{\label{fig:electrons_a}}
%    \subfigure{\label{fig:electrons_b}}
%\includegraphics[width=0.8\columnwidth]{Quinn}
%\includegraphics[width=0.8\columnwidth]{Parker}
%\caption{\label{fig:electrons}Double differential electron-deuteron cross section, $d^2\sigma/d\omega d\Omega$. The results obtained using the plane-wave approximation and the full calculation are compared with the experimental data reported by (a) Quinn \etal{}~\cite{Quinn:1988ua} and (b) Parker \etal{}~\cite{Parker:1986}. The panels are labeled according to the beam energy and scattering angle. Because Quinn \etal{} identified a $\sim$10\% normalization discrepancy in the dataset of panel (b), these data are also shown multiplied by 1.1.}
%\end{figure}

In Fig.~\ref{fig:electrons_a}, we compare our predictions with the double-differential cross section from Ref.~\cite{Quinn:1988ua}, presenting both the plane-wave result, which only includes the first term in Eq.~\eqref{eq:fullMatrixElement}, and the full calculation, which also accounts for the rescattering term. At the lowest energy transfers, $26\leq\omega\leq32$~MeV, the measurement probes the kinematics corresponding to Bjorken $x$ between 1.8 and 1.4, where electrons can elastically scatter off the deuteron. As this process is not accounted for in our framework, we underestimate the experimental results in this region. Otherwise, the full calculation reproduces the cross section reported by Quinn \etal{} with excellent accuracy, and the importance of accounting for the rescattering contribution is apparent.

A comparison with the measurement by Parker \etal{}~\cite{Parker:1986} shown in Fig.~\ref{fig:electrons_b} confirms the normalization discrepancy observed by Quinn \etal{}~\cite{Quinn:1988ua}. Should the normalization of this dataset be increased by 10\%, the results of Parker \etal{}~\cite{Parker:1986} would lead to similar conclusions as those of Quinn \etal{} For $\omega\leq55$~MeV, we underestimate the cross section corresponding to $1.4\leq x\leq1.9$, due to the missing elastic contribution. Beyond this region, the position and shape of the quasielastic peak are reproduced well by our full calculation.

% FIG. 5
\begin{figure}
\includegraphics[width=0.8\columnwidth]{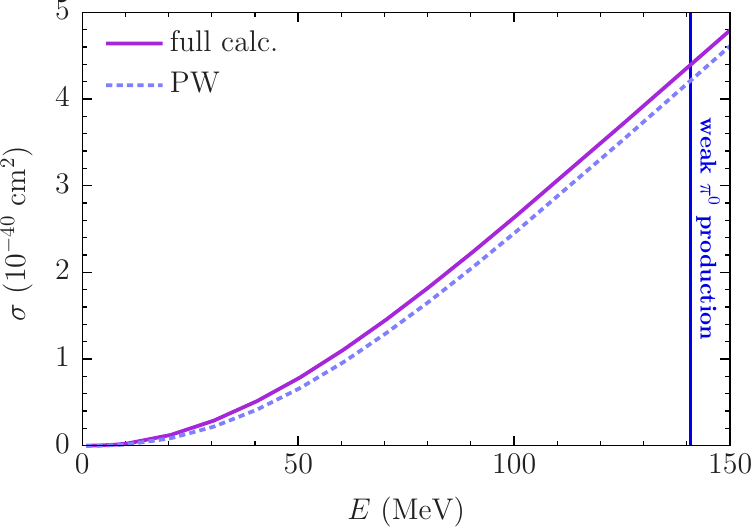}
\caption{\label{fig:total}Total cross section for positron capture on the deuteron as a function of the positron energy. Results obtained using the plane-wave approximation and the full calculations are presented. The vertical line indicates the threshold for the $e^+ + \isotope[2]{H} \rightarrow \bar\nu_e + p + p + \pi^0$ background channel. Note that a~fully longitudinally polarized beam is considered; for an unpolarized beam, the result should be halved. 
}
\end{figure}

% FIG. 6
\begin{figure}
\includegraphics[width=0.8\columnwidth]{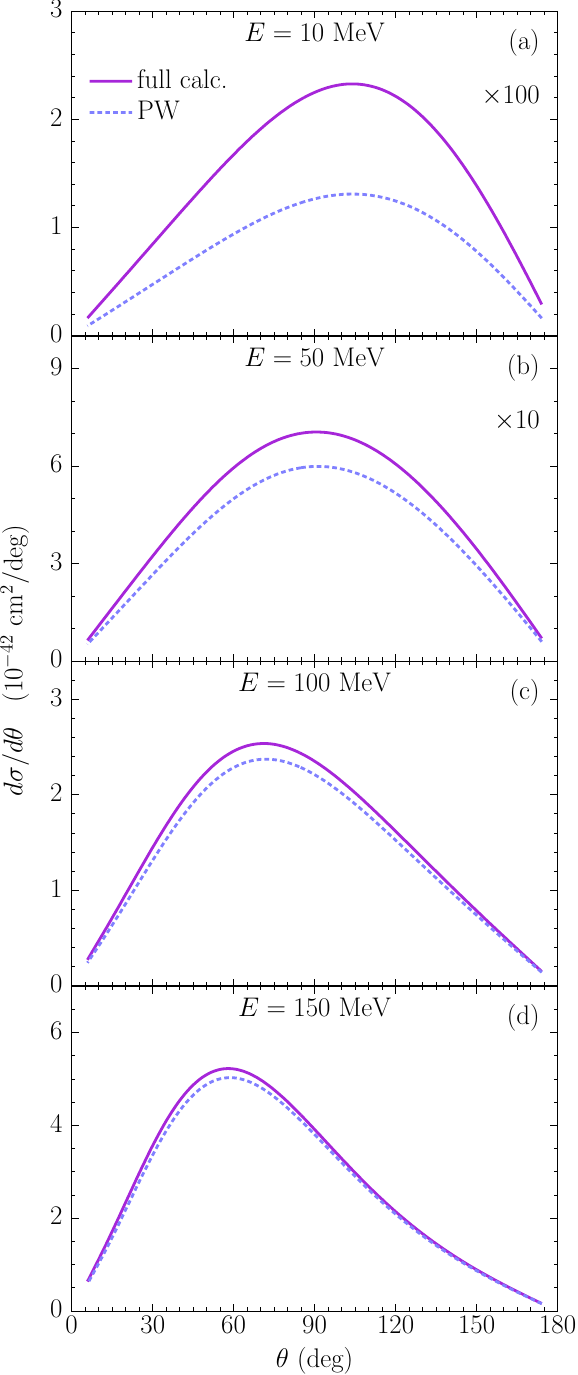}
\caption{\label{fig:dsdtheta}Single-differential cross section $d\sigma/d\theta$ as a function of the antineutrino scattering angle for positron capture on the deuteron, presented for various beam energies. Note that for visibility, the results in panels (a) and (b) are multiplied by 100 and 10, respectively.
}
\end{figure}

\begin{figure}[ht]
\includegraphics[width=0.8\columnwidth]{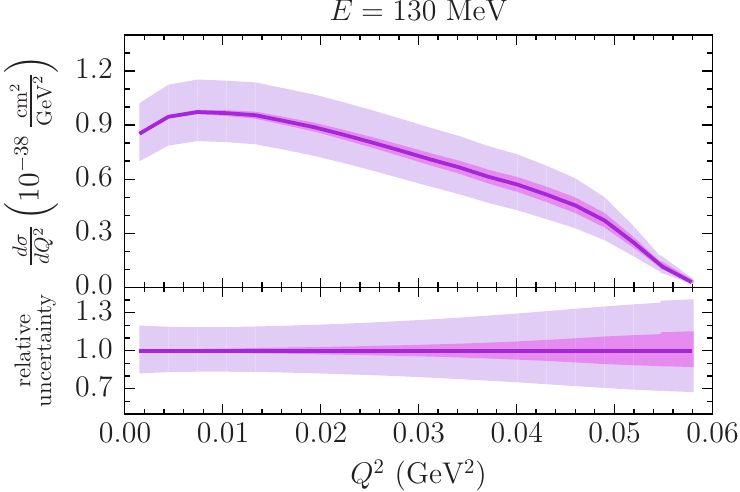}
\caption{\label{fig:130_dsdQ2}Single-differential cross section $d\sigma/dQ^2$ as a function of the four-momentum transfer squared for positron capture on the deuteron at beam energy 130~MeV. The upper panel presents the result together with the absolute uncertainty stemming from the axial form factor, and the lower panel shows the relative uncertainty. The inner (outer) bands correspond to the MINERvA parametrization (flat 10\%) uncertainty.
}
\end{figure}

\begin{figure}[h]
\includegraphics[width=0.8\columnwidth]{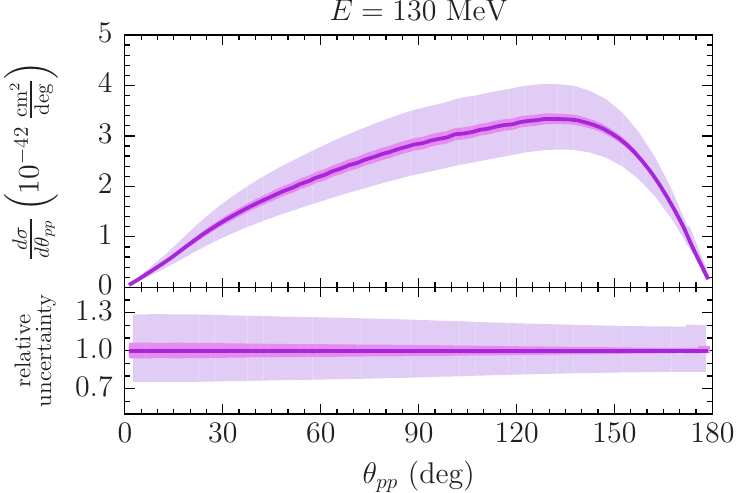}
\caption{\label{fig:130_dsdthetaPP}Same as Fig.~\ref{fig:130_dsdQ2} but for the single-differential cross section $d\sigma/d\theta_{pp}$ as a function of the angle between the final-state protons.
}
\end{figure}

\begin{figure}[h]
\includegraphics[width=0.8\columnwidth]{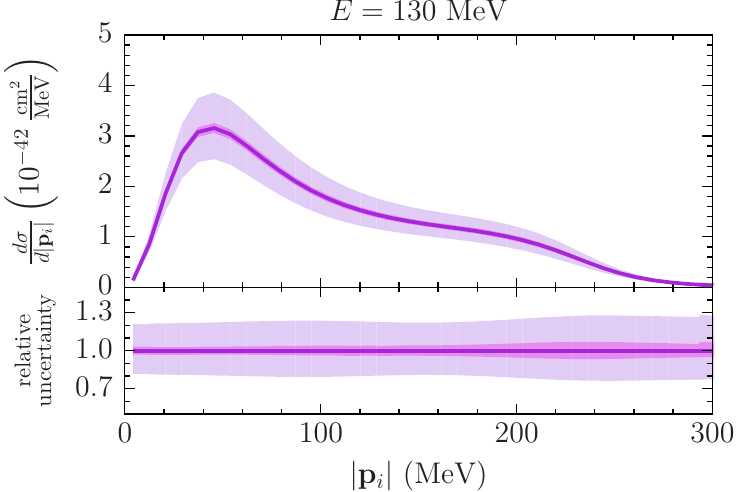}
\caption{\label{fig:130_dsdp}Same as Fig.~\ref{fig:130_dsdQ2} but for the single-differential cross section $d\sigma/d|\ve p_i|$ as a function of the momentum of a final-state proton.
}
\end{figure}

As explained in Sec.~\ref{sec:formalism}, in this article we do not account for the contribution of two-nucleon currents. Their role is known to increase when the beam energy or scattering angle increases~\cite{Marcucci11,PRC86.035503}. Therefore, the results for electron scattering---obtained at higher beam energies and higher scattering angles than those relevant to the main topic of our analysis---suggest that accounting only for single-nucleon currents is a good approximation at the considered kinematics.

Having verified the accuracy of our framework, we turn to its predictions for positron capture on the deuteron. In Fig.~\ref{fig:total}, we present the total cross section obtained using the full calculation and the plane-wave result. As the positron energy increases, the cross section grows quadratically up to $\sim$80~MeV, above which the growth becomes linear. The importance of the rescattering contribution, neglected in the plane-wave prediction, quickly diminishes with energy.

To illustrate this point clearly, in Fig.~\ref{fig:dsdtheta} we compare the full and plane-wave calculations for the differential cross section $d\sigma/d\theta$ as a function of the antineutrino scattering angle, obtained for positron energies between 10 and 150~MeV. While not directly observable, the value of $\theta$ can be reconstructed from the equation
\[
\cos\theta = \frac{\n{k}^2 + E'^2 - \n{p_1}^2 -2\n{p_1}\n{p_2}\cos\theta_{pp} - \n{p_2}^2}{2\n{k}E'},
\]
where the angle between the final-state protons' momenta is denoted as $\theta_{pp}$, the positron's momentum is $\n{k}=\sqrt{E^2-m_e^2}$, and the antineutrino energy is
\[
E'= E + M_d - \sqrt{M_p^2+\n{p_1}^2} - \sqrt{M_p^2+\n{p_2}^2}.
\]
As a consequence of energy and momentum conservation~\eqref{eq:energy&momentumConservation}, the antineutrino kinematics can be fully specified from the observables $\n{p_1}$, $\n{p_2}$, and $\theta_{pp}$.

In what follows, we confine our discussion to the full calculation for $E = 130$~MeV, where a measurement of the axial form factor is safe from pion backgrounds. We consider two scenarios for $F_A$ uncertainties: the estimate from the MINERvA parametrization~\cite{MINERvA:2023avz}, and a flat 10\% uncertainty. The second scenario is particularly well suited for evaluating the experimental sensitivity of differential cross sections to $F_A$ variations and assessing their potential to constrain the form factor in a model-independent manner. 

We estimate that at this positron energy, the total cross section is $3.92$, with uncertainties of $^{+0.16}_{-0.15}$ derived from the MINERvA $F_A$ parametrization and $^{+0.91}_{-0.79}$ under the flat 10\% uncertainty scenario, all in units of $10^{-40}$~cm$^2$. The effect of the $F_A$ uncertainties on various differential cross sections is examined in Figs.~\ref{fig:130_dsdQ2}--\ref{fig:130_dsdp}.

As shown in Fig.~\ref{fig:130_dsdQ2}, the differential cross section $d\sigma/dQ^2$ in the MINERvA parametrization scenario exhibits an uncertainty exceeding 5\% for $Q^2 \geq 0.032$~GeV$^2$, which rises to over 10\% for $Q^2 \geq 0.047$~GeV$^2$. Under the flat uncertainty scenario, the corresponding uncertainty exceeds 20\% for $Q^2 \geq 0.027$~GeV$^2$ and reaches 30\% or more for $Q^2 \geq 0.050$~GeV$^2$.

Figure~\ref{fig:130_dsdthetaPP} illustrates that the differential cross section $d\sigma/d\theta_{pp}$ as a function of the angle between the momenta of the final-state protons exhibits higher sensitivity to $F_A$ uncertainties at smaller angles. Specifically, the uncertainty exceeds 5\% for $\theta_{pp} \leq \ang{60}$ under the MINERvA parametrization scenario, and surpasses 20\% for $\theta_{pp} \leq \ang{105}$ in the flat uncertainty scenario. This behavior can be readily understood: because the nucleons forming the deuteron have their momenta in a back-to-back orientation, the largest momentum transfers correspond to the lowest values of~$\theta_{pp}$.

The differential cross section $d\sigma/d|\ve p_i|$ ($i=1, 2$) with respect to the momentum of a final-state proton has an uncertainty of 5--7\% for momenta between 190 and 300~MeV under the MINERvA parametrization scenario, which increases to 21\% or more for momenta exceeding 195~MeV under the flat uncertainty scenario; see Fig.~\ref{fig:130_dsdp}. This high-momentum region receives contributions from the largest momentum transfers.

\begin{figure}
\includegraphics[width=0.8\columnwidth]{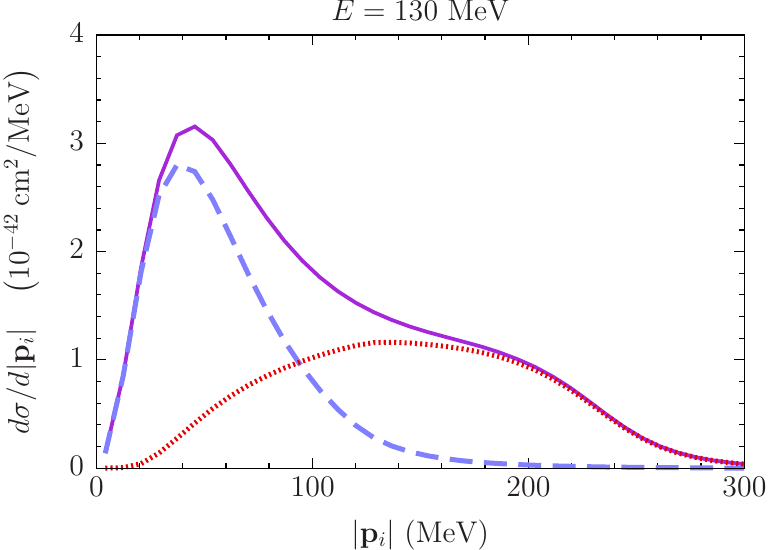}
\caption{\label{fig:130_dsdp_contribs}Single-differential cross section $d\sigma/d|\ve p_i|$ as a function of the momentum of a final-state proton for positron capture on the deuteron at beam energy 130 MeV. The dashed (dotted) line shows the lower-momentum (higher-momentum) contribution, and the solid line represents the total result. 
}
\end{figure}

\begin{figure}
\includegraphics[width=0.8\columnwidth]{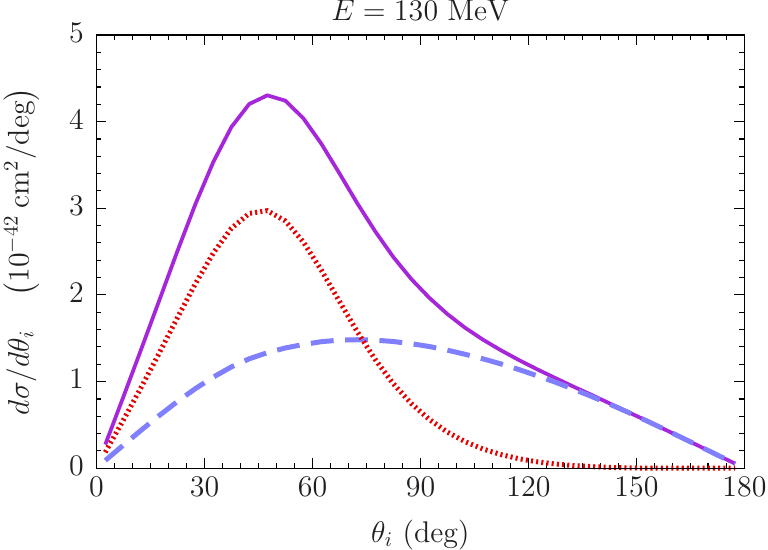}
\caption{\label{fig:130_dsdthetaP_contribs}Same as Fig.~\ref{fig:130_dsdp_contribs} but for the single-differential cross section $d\sigma/d\theta_i$ as a function of the emission angle of a final-state proton.
}
\end{figure}

%In what follows, we confine our discussion to the full calculation for $E = 130$~MeV, where a measurement of the axial form factor is safe from pion backgrounds. We estimate that at this positron energy, the total cross section is $(3.92\pm0.15)\times10^{-40}$~cm$^2$, where the uncertainty stems from the uncertainties in the $F_A$ parametrization~\cite{MINERvA:2023avz}. Their effect on various differential cross sections is examined in Figs.~\ref{fig:130_dsdQ2}--\ref{fig:130_dsdp}.
%
%As shown in Fig.~\ref{fig:130_dsdQ2}, the differential cross section $d\sigma/dQ^2$ has an uncertainty exceeding 5\% for $Q^2 \geq 0.032$ GeV$^2$, rising to 10\% or more for $Q^2 \geq 0.047$~GeV$^2$.
%
%Figure~\ref{fig:130_dsdthetaPP} illustrates that the differential cross section with respect to the angle between the momenta of the final-state protons, $d\sigma/d\theta_{pp}$, exhibits the largest sensitivity to the $F_A$ uncertainties---exceeding 5\%---when $\theta_{pp} \leq 60\degree$. This behavior can be readily understood: as the nucleons forming the deuteron have their momenta in a back-to-back orientation, the largest momentum transfers correspond to the lowest values of $\theta_{pp}$.
%
%The differential cross section with respect to the momentum of a final-state proton, $d\sigma/d|\ve p_i|$ ($i=1, 2$), has a 5--7\% uncertainty for momenta between 185 and 300 MeV; see Fig.~\ref{fig:130_dsdp}. This region receives contributions from the largest momentum transfers.

To address realistic experimental conditions, we now examine how the cross-section measurement is affected by the proton detection threshold and the angular acceptance of the detector. In Fig.~\ref{fig:130_dsdp_contribs}, we decompose $d\sigma/d|\ve p_i|$ into the higher- and lower-momentum contributions, whose strengths are equal by construction. One needs to bear in mind that the kinematics of the spectator nucleons---which are typically slower than the struck nucleons---are largely independent of the momentum transfer. Because identifying a positron-capture event requires detecting both protons in coincidence, an event is lost if the lower-momentum proton falls below the threshold. A cutoff at a given momentum thus removes not only the sub-threshold protons themselves, but also their higher-momentum coincidence partners, effectively doubling the impact of the lower-momentum contribution below that cutoff. For example, a threshold of 10~MeV (20~MeV) [50~MeV] makes 3\% (11\%) [52\%] of coincidence events undetectable. 

In the absence of final-state interactions, the spectator nucleons would have a random angular distribution. To a large extent, this is true for the lower-momentum contribution, as shown in Fig.~\ref{fig:130_dsdthetaP_contribs}. Therefore, detecting coincidence events requires broad angular acceptance. A coverage up to $\ang{125}$~\cite{Burkert:2020akg} leaves only 16\% of events undetectable in the backward hemisphere.

\section{Summary and conclusions}
\label{sec:summary}
In this article, we considered the weak process of positron capture on the deuteron at beam energies below pion production thresholds. Employing a framework validated against electron-scattering data, we calculated the total cross section as a function of energy. We performed a detailed analysis of various differential cross sections at 130~MeV, focusing on their sensitivity to the nucleon axial form factor. We also discussed how detection thresholds and angular acceptance affect the measurement.

Our results demonstrate that in the kinematic regime where the state-of-the-art $F_A$ measurement~\cite{MINERvA:2023avz} has $\sim$60--100\% uncertainties (corresponding to the three points at the lowest $Q^2$ in Fig.~\ref{fig:MINERvA}), it is within reach of a positron-capture experiment to set new precision standards. 

Its potential impact is difficult to overestimate: enabling tests of lattice QCD predictions, reducing uncertainties in modeling neutrino interactions for precision oscillation studies, and opening prospects for positron-capture measurements at high $Q^2$, where mitigating pion production backgrounds can be challenging. 
 
%The positron upgrade at JLab would make Hall B an ideal place to perform such an experimental study, bringing the idea of Hwang~\cite{Hwang:1986nz} to fruition some four decades after its original proposal.

The phase-I of the positron upgrade at JLab presents an immediate opportunity to perform this measurement at the injector, bringing the idea of Hwang~\cite{Hwang:1986nz} to fruition some four decades after its original proposal.

\begin{acknowledgments}
A.M.A. would like to express his gratitude to Tejin Cai for the enlightening correspondence on the axial form factor extraction in MINERvA. 
This work was supported by the National Science Centre, 
Poland under Grant IMPRESS-U 2024/06/Y/ST2/00135 and
in part by the Excellence Initiative---Research University Program 
at the Jagiellonian University in Krak\'{o}w. E.P.R.'s contribution was supported by the National Science Centre, Poland under Grant No. 2020/38/E/ST2/00112. W.P's contribution was supported by NSF EAGER: \mbox{IMPRESS-U} award \#2427848. H.K’s contribution was supported by a Grant-in-Aid for Scientific Research (Grant No. JP25K07301) from the Japanese Ministry of Education, Culture, Sports, Science and Technology (MEXT).
The numerical calculations were
partially performed on the supercomputer cluster of the JSC, J\"ulich, Germany.
\end{acknowledgments}

\end{document}